\documentstyle[12pt,epsfig]{article}
\begin{document}

\begin{center}
\large
New Limits to the IR Background: \\ Bounds on Radiative Neutrino Decay
and on VMO Contributions to the Dark Matter Problem
\end{center}
\normalsize
\vskip .5in

\newcommand{\oxford}{$^{1}$ }
\newcommand{\wash}{$^{2}$ }
\newcommand{\leeds}{$^{3}$ }
\newcommand{\dublin}{$^{4}$ }
\newcommand{\iowa}{$^{5}$ }
\newcommand{\purdue}{$^{6}$ }
\newcommand{\caltech}{$^{7}$ }
\newcommand{\whip}{$^{8}$ }
\begin{center}
S.D.~Biller,\oxford J.~Buckley,\wash 
A.~Burdett,\leeds J.~Bussons~Gordo,\dublin D.A.~Carter-Lewis,\iowa 
D.J.~Fegan,\dublin J.~Finley,\purdue J.A.~Gaidos,\purdue
A.M.~Hillas,\leeds  F.~Krennrich,\iowa R.C.~Lamb,\caltech
R.~Lessard,\purdue J.E.~McEnery,\dublin G.~Mohanty,\iowa 
J.~Quinn,\dublin A.J.~Rodgers,\leeds H.J.~Rose,\leeds 
F.~Samuelson,\iowa G.~Sembroski,\purdue P.~Skelton,\leeds T.C.~Weekes,\whip 
J.~Zweerink\iowa
\\[2ex]
{\em \oxford Oxford University, Oxford (UK)\\
\wash Washington University, St. Louis, Mo (USA)\\
\leeds University of Leeds, Leeds (UK)\\
\dublin University College, Dublin (Ireland)\\
\iowa Iowa State University, Ames, IA (USA) \\
\purdue Purdue University, West Lafayette, IN (USA)\\
\caltech California Institute of Technology, California(USA) \\
\whip Whipple Observatory, Amado, AZ (USA)\\}
\end{center}

\vskip .5in

\begin{abstract}
\noindent  From considering the effect of $\gamma$-$\gamma$
interactions on recently observed TeV gamma-ray spectra, improved
limits are set to the density of extragalactic infrared (IR) photons 
which are robust and essentially model-independent.
The resulting limits are up to two orders of magnitude more
restrictive than direct observations in the 0.025-0.3 eV regime.
These limits are used to improve constraints on radiative neutrino decay in the
mass range above 0.05eV and on Very Massive Objects (VMOs) as
providing the dark matter needed to explain galaxy rotation curves.
\end{abstract}

\vskip 0.5in
\begin{center}
{\large \em To Appear In Physical Review Letters}
\end{center}

\newpage

\section{Introduction}

	The extragalactic background IR field
potentially contains a wealth of information relevant to both cosmology 
and particle physics (see for example \cite{Bond}),
but has so far eluded any conclusive detection. Direct measurement
of the characteristics of this radiation are frustrated by the
dominance of local, galactic IR sources.
TeV gamma-ray astronomy provides a means to 
study the IR background indirectly, free of such complications,
by looking for modifications to high energy gamma ray spectra
due to interactions with this background field via the process 
$\gamma \gamma \longrightarrow e^{+} e^{-}$.
This idea was first noted by Gould and Schr\'{e}der \cite{Gould} and 
has been more recently restated by Stecker and de Jager \cite{Stecker}.

	Previous IR studies making use of this phenomenon have generally used 
spectral data from the active galactic nucleus (AGN) Mrk 421 
taken by the EGRET \cite{Hartman} 
and Whipple \cite{Mohanty} experiments at GeV and TeV energies, respectively.
These studies have relied on extrapolating flux measurements between these 
two energy regimes in an effort to determine the inherent shape of
the source spectrum \cite{Stecker} \cite{Dwek} \cite{DeJager} \cite{Biller}. 
This approach has several difficulties:
{\bf 1)} Since the flux level is known to be highly variable at TeV
energies, the relative 
normalization between EGRET and Whipple flux measurements is not
known; {\bf 2)} The flux of IR photons cannot be assumed to be
zero in the intervening energy range; and
{\bf 3)} The shape of the inherent source spectrum is not known
and cannot be assumed to necessarily follow a single power law
from GeV to TeV energies. Indeed, current best fit values for Mrk 421
at TeV energies indicate a steeper spectrum than is reflected
by the EGRET data \cite{Zweerink}, suggesting that the spectral shape changes
in the intervening energy range. In addition, most of the analyses have 
assumed a particular model to describe the shape of the IR photon 
spectrum. Even a more recent attempt at a less model-dependent limit using 
data from Mrk 501 nevertheless assumes a power-law source spectrum, 
an arbitrary IR flux normalization and constant IR energy densities over 
energy ranges spanning up to an order of magnitude in extent \cite{Stanev}.
Limits derived in this manner may not apply 
if the background IR spectrum is of a different shape and, in particular, 
cannot necessarily be used to constrain mechanisms which might produce sharp 
features in the spectrum on a scale smaller than that of the assumed model.
This current paper will attempt to define more robust limits based on recent 
high energy observations of AGN which do not rely on extrapolation 
from lower energies or the assumption of a strict power-law source spectrum
and are independent of the IR model down to scales corresponding to
a factor of 2 change in IR energy. These limits will then be used to constrain 
significantly the possible contribution of Very Massive Objects (VMOs) to the 
dark matter problem and also to place improved bounds on the 
radiative decay of massive neutrinos in the mass range above $\sim0.05$ eV, 
which is in the regime of interest to the atmospheric neutrino problem.

\section{Derived Upper Limits to the Background IR}

	Two active galaxies, Mrk 421 and Mrk 501, have thus far been seen 
to produce TeV gamma-ray emission by independent experimental groups.
These active galaxies are both of the blazar class
and are located at redshifts of 0.031 and 0.033, respectively.
Both objects have exhibited significant flux variability at
TeV energies, changing by an order of magnitude or more
with characteristic timescales of less than an hour 
(in the case of Mrk 421, as short as 15 minutes). Recent
analyses of data from Mrk 421 by the Whipple group 
\cite{Krennrich} \cite{Zweerink} indicate
a differential spectrum with an $\sim E^{-2.6}$ power law extending
to energies beyond 5 TeV, possibly as high as 8 TeV.
Similarly, the HEGRA collaboration has measured the ``high-state''
spectrum of Mrk 501 \cite{HEGRA} and has found it to be described by an
$\sim E^{-2.5}$ power law extending to energies as
high as 10 TeV.

	For the purposes of this paper, the available data on active
galaxies at a redshift of $\sim0.03$ will thus be taken to indicate a 
spectral shape which is bounded at the 90\% confidence level between the 
power laws $E^{-2.3}$ and $E^{-2.8}$ and normalized by an arbitrary flux 
constant. It will be assumed that the measured spectra continue up to 
energies of 10 TeV. However, due to possible systematic uncertainties 
in energy scale of $\sim$30\%, more conservative limits will also be 
computed assuming a maximum gamma-ray energy of 6 TeV.
Finally, it will be assumed that the inherent differential source spectrum
is steeper than an $E^{-2}$ power law, although the 
specific shape need not be a power law itself.
The limits derived are relatively insensitive to this latter choice.
For example selecting $E^{-1.5}$ as a bound was found to change the 
inferred limits by less than 30\%.

	For ease of computation, the high energy spectrum is taken to
consist of differential flux measurements at the following discrete 
energies: 0.4, 0.75, 1.0, 1.5, 2.0, 3.0, 4.0, 6.0 and 10.0 TeV. 
The spectrum of IR photons is approximated by a series of separate 
intervals spanning the energies 0.025-0.042, 0.042-0.083, 0.083-0.167, 
0.167-0.33, 0.33-0.625 and 0.625-1.0 eV. Within each interval, the value 
of $\varepsilon^2 n(\varepsilon)$ is taken as constant, where $\varepsilon$ 
is the energy of the IR photon and $n$ is the number density of photons 
({\em i.e.} photons/$cm^3$/eV). The value of $\varepsilon^2 n(\varepsilon)$ 
among these intervals can thus be varied to reflect different background IR 
models, and the corresponding effect of $\gamma$-$\gamma$ absorption on 
TeV gamma-rays calculated according to the prescription 
described by previous papers \cite{Stecker}, \cite{Biller}. For these 
calculations, a conservative value for the Hubble constant of 85 km/s/Mpc 
is assumed \cite{PDG}. 

	To calculate upper limits then, the inherent AGN differential 
spectrum was allowed to assume any shape steeper than $E^{-2}$ and
the values of $\varepsilon^2 n(\varepsilon)$ among IR intervals were 
allowed to take on any values 
subject to the following constraints: {\bf 1)} that these values are
lower than the directly determined upper bounds imposed by other 
experiments (shown in figure 1); and {\bf 2)} that the attenuation implied
by the IR model would not cause the shape of the observed
TeV spectrum to deviate outside of the power-law bounds previously defined
based on the data. This approach essentially anchors the lower energy TeV
data to the appropriate observational upper bounds and then extends these
bounds based on the shape and extent of the AGN spectra 
at higher energies. From these considerations, the maximum value of energy
density permitted in each IR interval was determined.
The resulting limits are summarized in table 1 and are also plotted in 
figure 1 for assumed maximum gamma-ray energies of 6 TeV and 10 TeV. 
Note that this choice of maximum energy primarily effects only the 
two lowest energy IR intervals.

\begin{table}[h]
\begin{center}
\begin{tabular}{|l|l|l|}
\hline
 & \multicolumn{2}{c|}{Upper Limit to $\varepsilon^2 n(\varepsilon)$ 
(eV/cm$^3$)} \\
Energy Interval (eV) & $E_{max}$ = 10 TeV & $E_{max}$ = 6 TeV \\
\hline
0.025-0.042   & 0.0064 &        \\
0.042-0.083   & 0.0027 & 0.094  \\
0.083-0.167   & 0.0027 & 0.0038 \\
0.167-0.33    & 0.0035 & 0.0037 \\
0.33-0.625    & 0.0052 & 0.0052 \\
0.625-1.0     & 0.012  & 0.012 \\
\hline
\end{tabular}
\end{center}
\caption{Derived upper limits to $\varepsilon^2 n(\varepsilon)$}
\end{table}

	In the range of $0.025-0.33$ eV, the limits 
presented here are up to 2 orders of magnitude more restrictive 
than those of DIRBE \cite{Hauser}. Note that these limits apply only to 
models which would produce features in the IR spectrum on a scale 
comparable to or larger than the interval widths used in the calculation. 
Monoenergetic line features, for example, are not necessarily constrained 
by these limits. 

\section{Constraints on VMO Abundance}

	Under certain cosmological scenarios, density fluctuations in
the early universe at a redshift of $\sim$1000 could give rise to the
pregalactic formation of massive stars ($200<M/M_{\odot}<10^5$) at 
redshifts in the range of $\sim$100-300 \cite{BAC}. Such Very Massive Objects
(VMOs) would collapse to black holes after their main-sequence phase
without any significant metal ejection. It has been suggested that
the remnants of VMOs, which would ultimately cluster with galaxies, might 
provide an explanation for the dark matter associated with galaxy rotation
curves \cite{Carr}, \cite{Faber}. Such rotation curves indicate that the
amount of dark matter is in excess of 5 percent of the critical density,
while other studies of galaxy motion and hot gas in clusters of galaxies
suggest a value as high as 20 percent \cite{PDG}.

While invisible now, VMOs would
have produced a significant flux of IR photons during their burning phase.
The peak number density of these photons is given \cite{Bond} by:

\[\varepsilon_{pk}^2 n_V(\varepsilon_{pk})\simeq40\frac{\Omega_V h^2}{1+z_V}\]

\[\varepsilon_{pk} \simeq \frac{30}{1+z_V}\]

\noindent where $\varepsilon_{pk}$ is the IR photon energy (in eV) 
corresponding to the peak of the VMO contribution,
$n_V$ is the photon number density due to VMOs in 
$cm^{-3} eV^{-1}$, $\Omega_V$ is the mass-density of VMOs expressed as 
a fraction of the critical density, $z_V$ is the formation redshift of the 
VMOs, and the Hubble constant is given by $H_0=100h$ km/s/Mpc.
Thus, the IR limits derived previously may be used to place upper
bounds on $\Omega_V$. These bounds are shown in figure 2 as a function
of the assumed VMO formation redshift for $h$=0.65 (which is conservative for
this limit). Note that for either maximum gamma-ray energy assumed, the 
limit on $\Omega_V$ is less than 0.05 for
the redshift range of 100-200 and less than 0.065 for the range 200-300.
If the maximum gamma-ray energy is taken to be 10 TeV, the limit is
less than 0.05 for the the entire 100-300 range.
Hence, assuming minimal reprocessing by dust grains, these limits appear to 
rule out VMO models as providing the explanation for observed galaxy rotation
curves.

\section{Constraints on Radiative Neutrino Decay}

	Several experiments have observed an apparent descrepency in
the ratio of $\nu_e$ to $\nu_{\mu}$ fluxes expected from cosmic ray
interactions in the atmosphere \cite{IMB} \cite{Kamiokande} \cite{Soudan}.
A possible explanation for this effect may involve 2-component neutrino 
oscillations with $\Delta m^2 \sim 0.01$eV$^2$ and, thus, a neutrino mass in 
the regime of 0.1 eV (provided the mass of the lighter neutrino is 
comparable to or smaller than this value).
If neutrinos have mass and different flavor eigenstates may
mix, then massive neutrinos might decay via the process 
$\nu_H \rightarrow \nu_L + \gamma$, where $m_{\nu_H} > m_{\nu_L}$.
For neutrino masses in the $\sim$eV regime, such a radiative decay process
would contribute to the infrared background. Although mono-energetic
photons would be initially produced from a 2-body decay, the photon
spectrum would be broadened for relic neutrinos which decay over a 
range of redshifts such that the background density can be constrained
by the limits presented previously.

	Some unstable relic neutrinos with a lifetime longer than that of
the universe would still be decaying now and would have produced a 
background photon density given by:
\[ n_{\gamma}(\epsilon) \simeq 
1.1\times10^{12} \frac{B_r}{\tau_{\nu} h \epsilon} \]
where $\epsilon$ is half the rest mass of the unstable neutrino in eV,
$n_{\gamma}$ is the photon number density due to neutrino decay 
in $cm^{-3} eV^{-1}$,
$B_r$ is the branching ratio for the radiative decay mode and $\tau_{\nu}$
is the neutrino lifetime in years \cite{Ressell}.
Implied lower bounds for $\tau_{\nu}/B_r$  in the neutrino mass range
$0.1eV < m_{\nu} < 1eV$ are shown in figure 3 based on the IR limits
presented in this paper for $h=0.85$. Previous constraints in this mass range 
from the work of Ressell and Turner \cite{Ressell} are shown for comparison. 
This represents an improvement of the constraints in some cases by more 
than 2 orders of magnitude.

	Similarly, neutrinos with lifetimes shorter than the age of 
the universe but longer than the age at which decoupling of matter 
and radiation occurred will result in a background photon density
given by:
\[n_{\gamma}(\epsilon) \simeq 115 \frac{B_r}{\epsilon} \]
where, in this case,
\[\epsilon=\left(\frac{\frac{1}{2}m_{\nu}c^2}{eV}\right) 
\left(\frac{\tau_{\nu}\sqrt{\Omega_o}h}{6.5\times10^9yr}\right)^{2/3} \]
where $\Omega_o$ is the mass-density of the universe in units of the
critical density.
IR limits given here have thus been used to define exclusion contours
which can be conveniently characterised by
\[ B_r < 10^{-4}\left(\frac{1 eV}{m_o}\right) \]
for the region defined by
\[m_{\nu}=\left(\frac{m_o}{1 eV}\right)
\left(\frac{6.5\times10^9yr}{\tau_{\nu}\sqrt{\Omega_o}h}\right)^{2/3}\]
where $0.05 eV <m_o<2.0 eV$ for an assumed maximum gamma-ray energy of 10 TeV, 
and $0.1 eV <m_o<2.0 eV$ for a maximum energy of 6 TeV. Once again, this represents 
an improvement in branching ratio sensitivity of approximately two orders of magnitude.

\section{Discussion}

	Recent data from very high energy gamma-ray observations
of active galaxies has been used to set improved upper limits to
the infared background radiation which are robust and relatively 
model-independent. These limits cast doubt on VMO models as providing 
the explanation for dark matter in galactic halos and tighten
constraints on the radiative decay of relic neutrinos in the mass 
range near 0.1 eV by approximately two orders of magnitude. 

	Substantial improvement in the ability to study the infrared 
background via $\gamma$-$\gamma$ interactions is 
expected with the next generation of instruments which will allow
the behavior of more distant sources to be examined. In fact, 
the procedure used to derive the limits given in Table 1
may also be used to determine lower limits to the maximum distance
out to which TeV gamma-ray telescopes can be expected to probe.
This has been done for the gamma-ray energy range of 0.4-10 TeV,
assuming that a source remains ``visible'' out to an optical
depth of 2 and that the majority of the IR was produced prior to
the epoch of the sources under study.
IR and UV upper bounds from direct observations in the energy
range greater than 2 eV have also been used to calculate this
minimum distance for gamma-ray energies of 0.05 and 0.1 TeV.
The results are shown in figure 4. The current generation of
ground-based, gamma-ray telescopes have a typical lower energy threshold
of $\sim$0.5 TeV and are thus expected to be able to see sources possessing 
redshifts up to or beyond $z$=0.1. The next generation of instruments
is expected to have an energy threshold in the region of 0.05 to 0.1 TeV,
and will therefore be able to see out to a redshift of at least $z$=0.5.
This represents a substantially larger volume of space than has so far
been explored at TeV energies. This, and the known preponderance of AGN
at larger redshifts, indicates a bright future for ground-based, 
gamma-ray astronomy and strongly suggests a drive towards higher duty cycles,
larger apertures and greater dynamic ranges for such instruments to 
explore this regime thoroughly.

	This work has been supported in part by PPARC, Forbairt,
the US Department of Energy, NASA and the Smithsonian Institution.

\begin{figure}[p]
\setlength{\epsfxsize}{0.85\textwidth}
\leavevmode
\epsfbox{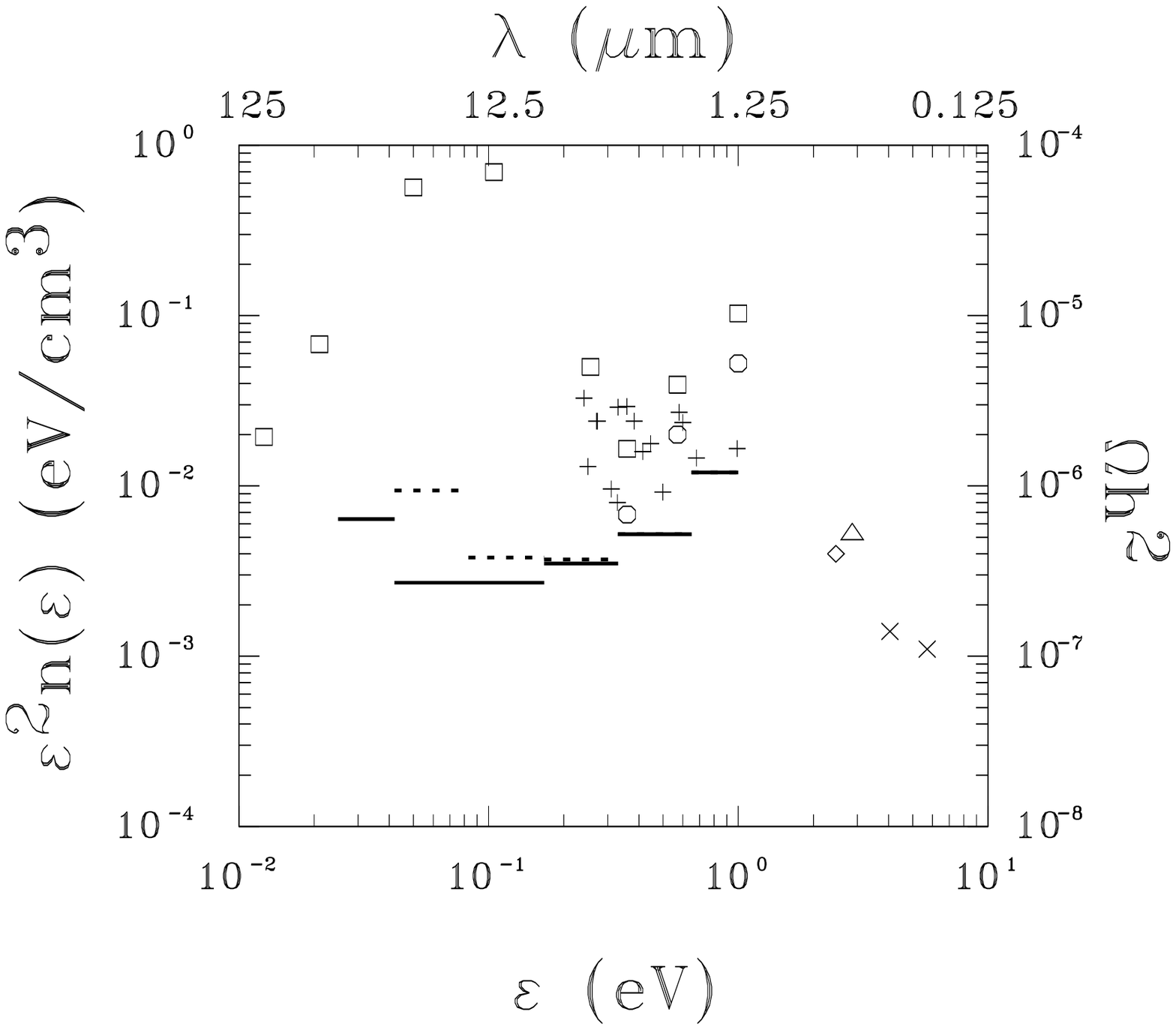}
\caption[Upper limits to the background]
{Upper limits to the background infrared background: 
``+'' symbols are from Matsumoto {\em et al.} \cite{Matsumoto};
the diamond is from Dube {\em et al.} \cite{Dube}; the triangle is from 
Toller \cite{Toller}; squares are from Hauser {\em et al.} \cite{Hauser}; 
circles are from Kashlinsky {\em et al.} \cite{Kashlinsky}; crosses are from 
Maucherat-Joubert {\em et al.} \cite{Maucherat}; the 
solid lines show the limits derived in this work for a maximum gamma-ray
energy of 10 TeV, while the dashed lines correspond to a maximum energy
of 6 TeV.}
\label{irlimits} \end{figure}

\begin{figure}[p] 
\setlength{\epsfxsize}{0.85\textwidth}
\leavevmode
\epsfbox{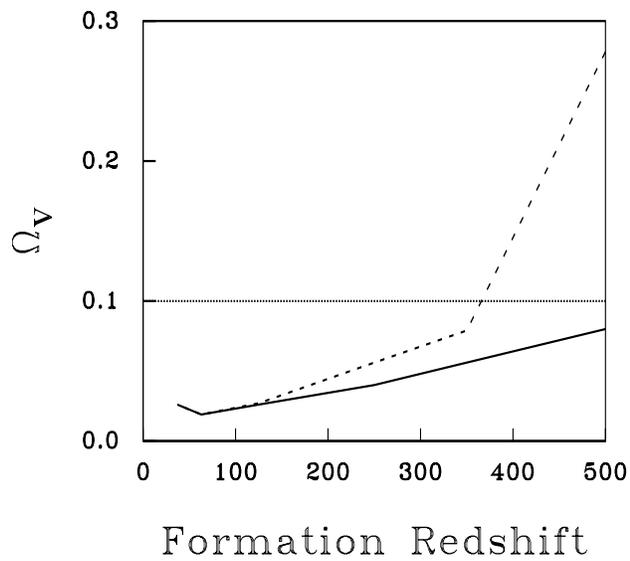}
\caption[Upper limits to the abundance
of Very Massive Objects]{Upper limits to the abundance
of Very Massive Objects based on bounds to the IR background
derived in this work. The limit is expressed as a fraction
of the critical density and is plotted as a function of the assumed
formation redshift of the VMOs. The solid line corresponds to an 
assumed maximum gamma-ray energy of 10 TeV and
the dashed line to a maximum energy of 6 TeV.}
\label{vmolim} \end{figure}

\begin{figure}[p]
\setlength{\epsfxsize}{0.85\textwidth}
\leavevmode
\epsfbox{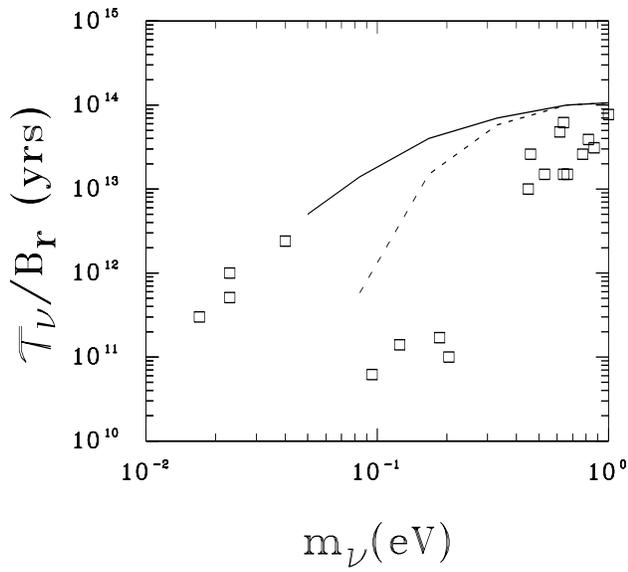}
\caption[Lower bounds to the ratio
of the lifetime to the branching ratio to radiative decay.]
{Lower bounds to the ratio of the neutrino lifetime to the branching 
ratio for radiative decay based on this current work for the case
where the neutrino lifetime is greater than the current age of 
the universe. The solid line shows the limits derived here assuming
a maximum gamma-ray energy of 10 TeV, while the dashed line coresponds
to a maximum energy of 6 TeV. Open squares show the limits derived 
previously by Ressell and Turner.}
\label{nulim} \end{figure}

\begin{figure}[p] 
\setlength{\epsfxsize}{0.85\textwidth}
\leavevmode
\epsfbox{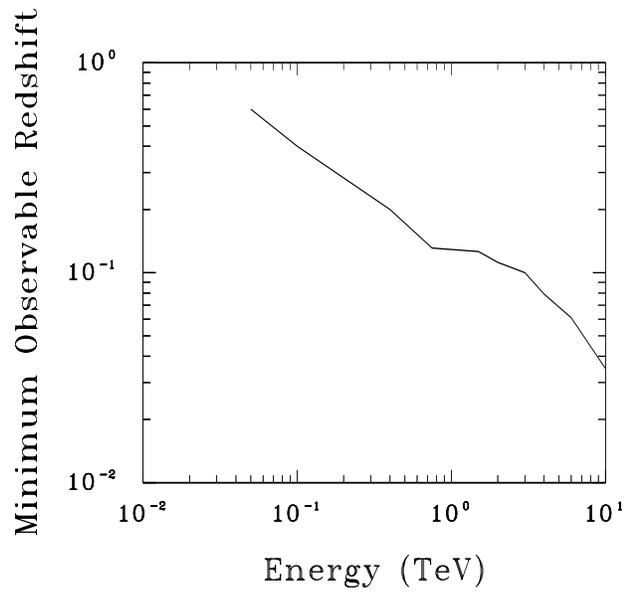}
\caption[Source visibility for TeV gamma-ray telescopes.]
{Lower limits to the maximum redshift out to which TeV gamma-ray sources are
visible as a function of primary gamma-ray energy, based on
limits to the background density of IR/UV photons.}
\label{visib} \end{figure}

\end{document}